\documentclass{article}
\usepackage{graphicx}
\usepackage{amsmath}

\input{tcilatex}

\begin{document}

\title{Size Information Obtained Using Static Light Scattering Technique}
\author{Yong Sun}
\maketitle

\begin{abstract}
Detailed investigation of static light scattering $\left( SLS\right) $ has
been attempted in this work using dilute water dispersions of homogenous
spherical particles, poly($N$-isopropylacrylamide) microgels and simulated
data. When Rayleigh-Gan-Debye approximation is valid, for large particles,
the simple size information, the static radius $R_{s}$ and distribution $%
G\left( R_{s}\right) $, can be accurately obtained from SLS. For small
particles, the root mean-square radius of gyration $\left\langle
R_{g}^{2}\right\rangle _{Zimm}^{1/2}$ and the molar mass of particles
measured using the Zimm plot are discussed. The results show that the molar
mass measured using the Zimm plot over the average molar mass of particles
is a function of the size distribution. With the assistance of simulated
data, the effects of the reflected light and noises have been investigated
in detail. Measuring the static radius from the SLS data provides one method
to avoid the stringent requirements for the sample quantity and the
instrument capability at small scattering angles.
\end{abstract}

\section{Introduction}

The intensity of the scattered light is determined by the sizes, shapes and
interaction among the particles in the scattering medium. During the last
few decades, dynamic light scattering $\left( DLS\right) $ is widely used to
obtain the size information of particles for colloidal dispersion systems.
Although the static light scattering $\left( SLS\right) $ spectroscopy
contains more sensitive size information, in general, the measurements of
SLS spectroscopy are simplified to\ the Zimm plot, Berry plot or Guinier
plot to obtain the root mean-square radius of gyration $\left\langle
R_{g}^{2}\right\rangle ^{1/2}$ and the molar mass of particles provided that
the particle sizes are small. Since it is hard to obtain the particle size
distribution for small poly-disperse particles using DLS technique, for
dilute poly-disperse homogeneous spherical particles, Pusey and van Megen 
\cite{re1} proposed a method to detect small poly-dispersities when the
Rayleigh-Gans-Debye $\left( RGD\right) $ approximation is valid, measuring
the dependence of the effective diffusion coefficient obtained from the
initial slope of the correlation function with respect to the scattering
angle. By definition, the effective diffusion coefficient is the
intensity-weighted average diffusion coefficient. Both theoretical and
experimental results show that the angular dependence of the effective
diffusion coefficient is a sensitive function of the particle size and
distribution.

How the particle size distributions can be obtained directly from the SLS
data has been researched by a few authors. Hallett and Strawbridge \cite{re2}
have studied the theoretical scattered intensity of a coated sphere with
vertically polarized incident light. Then the scattered intensity at the
geometrical or linear trial radii between $r_{\min }$ and $r_{\max }$ was
used to fit the SLS data. Schnablegger and Glatter \cite{re3} assumed that
the size distribution can be described as a series of cubic B-splines, then
used the simulated and measured data to demonstrate the computation
procedure.

In this article, we deal with the dilute poly-disperse homogeneous spherical
particles. We assume that the number distribution of particles is Gaussian
and we consider the effects of the form factor and the scattered
intensity-weighted differences of different size particles on the scattered
light intensity. Then with the assistance of a non-linear least squares
fitting program, the mean particle size $\left\langle R_{s}\right\rangle $
and the standard deviation $\sigma $ are obtained. With this treatment, we
can avoid the constraints of the Zimm plot, Berry plot or Guinier plot on
measurements, and their stringent dependences on sample quality and
instrument capability at small angles. For large particles, size
distributions can be measured accurately. With the assistance of simulated
data, the effects of the reflected light and noises have been investigated
in detail. Through the theoretical and simulated data analysis, the root
mean-square radius of gyration $\left\langle R_{g}^{2}\right\rangle
_{Zimm}^{1/2}$ and the molar mass of particles measured using the Zimm plot
are also discussed. The results show that the molar mass measured using the
Zimm plot over the average molar mass of particles is a function of the size
distribution. With theoretical and experimental data analysis, better
understanding of the size information contained in SLS spectroscopies is
obtained.

\section{Theory}

For simplicity, we consider homogeneous spherical particles and assume that
the RGD approximation is valid. The average scattered light intensity of a
dilute non-interacting polydisperse system in unit volume can be obtained
for vertically polarized light 
\begin{equation}
\frac{I_{s}}{I_{inc}}=\frac{4\pi ^{2}\sin ^{2}\theta _{1}n_{s}^{2}\left( 
\frac{dn}{dc}\right) _{c=0}^{2}c}{\lambda ^{4}r^{2}}\frac{4\pi \rho }{3}%
\frac{\int_{0}^{\infty }R_{s}^{6}P\left( q,R_{s}\right) G\left( R_{s}\right)
dR_{s}}{\int_{0}^{\infty }R_{s}^{3}G\left( R_{s}\right) dR_{s}},
\label{mainfit}
\end{equation}
where $\theta _{1}$ is the angle between the polarization of the incident
electric field and the propagation direction of the scattered field, $c$ is
the mass concentration of particles, $r$ is the distance between the
scattering particle and the point of the intensity measurement, $\rho $ is
the density of the particles, $I_{inc}$ is the incident light intensity, $%
I_{s}$ is the intensity of the scattered light that reaches the detector, $%
R_{s}$ is a static radius of a particle,$\ q=\frac{4\pi }{\lambda }n_{s}\sin 
\frac{\theta }{2}$ is the scattering vector, $\lambda $\ is the wavelength
of the incident light in vacuo, $n_{s}$\ is the solvent refractive index,$\
\theta $\ is the scattering angle, $P\left( q,R_{s}\right) $ is the form
factor of homogeneous spherical particles

\begin{equation}
P\left( q,R_{s}\right) =\frac{9}{q^{6}R_{s}^{6}}\left( \sin \left(
qR_{s}\right) -qR_{s}\cos \left( qR_{s}\right) \right) ^{2}  \label{P(qr)}
\end{equation}
and $G\left( R_{s}\right) $ is the number distribution. In this paper, the
number distribution is chosen as a Gaussian distribution

\begin{equation}
G\left( R_{s};\left\langle R_{s}\right\rangle ,\sigma \right) =\frac{1}{%
\sigma \sqrt{2\pi }}\exp \left( -\frac{1}{2}\left( \frac{R_{s}-\left\langle
R_{s}\right\rangle }{\sigma }\right) ^{2}\right) ,
\end{equation}
where $\left\langle R_{s}\right\rangle $ is the mean static radius and $%
\sigma $ is the standard deviation relative to the mean static radius.

If the reflected light is considered, the average scattered light intensity
in unit volume is written as 
\begin{equation}
\frac{I_{s}}{I_{inc}}=a\frac{4\pi \rho }{3}\frac{\int_{0}^{\infty
}R_{s}^{6}P\left( q,R_{s}\right) G\left( R_{s}\right)
dR_{s}+b\int_{0}^{\infty }R_{s}^{6}P\left( q^{\prime },R_{s}\right) G\left(
R_{s}\right) dR_{s}}{\int_{0}^{\infty }R_{s}^{3}G\left( R_{s}\right) dR_{s}}
\label{mainre}
\end{equation}
where

\begin{equation}
a=\frac{4\pi ^{2}\sin ^{2}\theta _{1}n_{s}^{2}\left( \frac{dn}{dc}\right)
_{c=0}^{2}c}{\lambda ^{4}r^{2}}
\end{equation}
and 
\begin{equation}
q^{\prime }=\frac{4\pi }{\lambda }n_{s}\sin \frac{\pi -\theta }{2}
\end{equation}
is the scattering vector of the reflected light. $b$ is a constant decided
by the shape of sample cell, the refractive indices of the solvent and the
sample cell and the geometry of instruments.

When the values of $qR_{s}$ are small, the form factor can be expanded and
Eq. \ref{mainfit} can be written as

\begin{equation}
\frac{4\pi ^{2}\sin ^{2}\theta _{1}n_{s}^{2}\left( \frac{dn}{dc}\right)
_{c=0}^{2}c}{\lambda ^{4}r^{2}N_{0}\frac{I_{s}}{I_{inc}}}=\frac{\left(
\int_{0}^{\infty }R_{s}^{3}G\left( R_{s}\right) dR_{s}\right) ^{2}}{%
\left\langle M\right\rangle \int_{0}^{\infty }R_{s}^{6}G\left( R_{s}\right)
dR_{s}}\left( 1+\frac{q^{2}\int_{0}^{\infty }R_{s}^{8}G\left( R_{s}\right)
dR_{s}}{5\int_{0}^{\infty }R_{s}^{6}G\left( R_{s}\right) dR_{s}}+\cdot \cdot
\cdot \right)
\end{equation}
where $N_{0}$ is the Avogadro's number, $\left\langle M\right\rangle $ is
the average molar mass of particles. It is defined as

\begin{equation}
\left\langle M\right\rangle =\frac{4\pi \rho N_{0}}{3}\int_{0}^{\infty
}R_{s}^{3}G\left( R_{s}\right) dR_{s}.
\end{equation}
Comparing with the Zimm plot analysis $\left[ \cite{re4}-\cite{re6}\right] $%
, the mean square radius of gyration, $\left\langle R_{g}^{2}\right\rangle
_{Zimm}$, for a polydisperse system is 
\begin{equation}
\left\langle R_{g}^{2}\right\rangle _{Zimm}=\frac{3\int_{0}^{\infty
}R_{s}^{8}G\left( R_{s}\right) dR_{s}}{5\int_{0}^{\infty }R_{s}^{6}G\left(
R_{s}\right) dR_{s}}  \label{RG}
\end{equation}
and the molar mass of particles measured using the Zimm plot is

\begin{equation}
M_{z}=\frac{\left\langle M\right\rangle \int_{0}^{\infty }R_{s}^{6}G\left(
R_{s}\right) dR_{s}}{\left( \int_{0}^{\infty }R_{s}^{3}G\left( R_{s}\right)
dR_{s}\right) ^{2}}.  \label{Momass}
\end{equation}

\section{Experiment}

The SLS spectroscopies were measured using the instrument built by ALV-Laser
Vertriebsgesellschaft m.b.H (Langen, Germany). It utilizes an ALV-5000
Multiple Tau Digital Correlator and a JDS Uniphase 1145P\ He-Ne laser to
provide a 23 mW vertically polarized laser\ at wavelength of 632.8 nm.

In this experiment, $N$-isopropylacrylamide (NIPAM, monomer) from Acros
Organics was recrystallized from hexane/acetone solution. Potassium
persulfate (KPS, initiator) and $N,N^{\prime }$-methylenebisacrylamide (BIS,
cross-linker) from Aldrich were used as received. Fresh de-ionized water
from a Milli-Q Plus water purification system (Millipore, Bedford, with a
0.2 $\mu m$ filter) was used throughout the whole experiment. The synthesis
of gel particles was described elsewhere $\left[ \cite{re7},\cite{re8}\right]
$ and the recipes of the\ batches used in this work are listed in Table 1.

\begin{center}
$\overset{\text{Table 1. Synthesis conditions for PNIPAM particles.}}{
\begin{tabular}{|c|c|c|c|c|c|}
\hline
Sample & $T\left( ^{o}C\right) $ & $t\left( hrs\right) $ & $%
W_{N}+W_{B}\left( g\right) $ & $KPS\left( mg\right) $ & $n_{B}/n_{N}$ \\ 
\hline
$PNIPAM-0$ & $70\pm 1$ & $4.0$ & $1.00$ & $40$ & $0$ \\ \hline
$PNIPAM-1$ & $70\pm 1$ & $4.0$ & $1.00$ & $40$ & $1.0\%$ \\ \hline
$PNIPAM-2$ & $70\pm 1$ & $4.0$ & $1.00$ & $40$ & $2.0\%$ \\ \hline
$PNIPAM-5$ & $70\pm 1$ & $4.0$ & $1.00$ & $40$ & $5.0\%$ \\ \hline
\end{tabular}
}$
\end{center}

The four samples were named according to the molar ratios $n_{B}/n_{N}$ of $%
N,N^{\prime }$-methylenebisacrylamide over $N$-isopropylacrylamide. They
were centrifuged at 14,500 RPM followed by decantation of the supernatants
and re-dispersion in fresh de-ionized water four times to remove of free
ions and any possible linear chains. Then the samples were diluted for light
scattering to weight factors of $5.9\times 10^{-6}$, $8.56\times 10^{-6}$, $%
9.99\times 10^{-6}$ and $8.38\times 10^{-6}$ \ for $PNIPAM-0$, $PNIPAM-1$, $%
PNIPAM-2$ and $PNIPAM-5$\ respectively. Before\ the measurements were made,
0.45 $\mu m$ filters (Millipore, Bedford) were used to do dust free for the
samples $PNIPAM-1,$ $PNIPAM-2$ and $PNIPAM-5$.

\section{Data Analysis}

How the size information is obtained from SLS is shown in this section. The
experimental data of the $PNIPAM$ microgel samples were used to show the
fitting process and the simulated data were used to examine the effects of
the different reflected light and the noises on the fit results and the
effects of the distribution on the molar mass of particles measured using
the Zimm plot.

\subsection{Experimental Data Analysis}

When Eq. \ref{mainfit} was fit to the data of $PNIPAM-1$ measured at a
temperature of 29$^{o}C$, it was found that the results for mean static
radii $\left\langle R_{s}\right\rangle $ and standard deviation $\sigma $
depended on the scattering vector range being fit, as shown in Table 4.1. If
a small scattering vector range is chosen, the parameters are not
well-determined. As the scattering vector range is increased, $\chi ^{2}$
and the uncertainties in the parameters decrease and $\left\langle
R_{s}\right\rangle $ and $\sigma $ stabilize. If the fitting scattering
vector range continues to increase, the values of $\left\langle
R_{s}\right\rangle $ and $\sigma $ begin to change and $\chi ^{2}$ grows.
This is the result of the deviation between the experimental and theoretical
scattered light intensity in the vicinity of the scattered intensity
minimum. This minimum lies at about the scattering vector 0.0177 $nm^{-1}$.
In this range, most of the scattered light is cancelled due to the light
interference. So many other characteristics of particles can show the
effects on the scattered light intensity, for example: the particle number
distribution deviates from a Gaussian distribution, the particle shape
deviates from a perfect sphere and the density of particles deviates from
homogeneity, etc. In order to avoid the effects of light interference, the
stable fit results during the scattering vector range from 0.00345 $nm^{-1}$
to 0.01517 $nm^{-1}$ are chosen as the size information obtained using the
SLS technique. In order to examine the effects of the different fitting
ranges of the scattering vector, the experimental data were fit again fixing
the larger value of $q$ and decreasing the fitting range. The fit results
also are shown in Table 4.1. The values show the fit results are
well-determined when the fitting range is enough large. Figure 4.1 shows the
fit results and the residuals during the scattering vector range from
0.00345 $nm^{-1}$ to 0.01517 $nm^{-1}$.

\begin{center}
$\overset{\text{Table 4.1 The fit results for }PNIPAM-1\text{ in different
scattering vector ranges and a temperature of 29}^{o}C\text{.}}{
\begin{tabular}{|c|c|c|c|}
\hline
$q\left( 10^{-3}nm^{-1}\right) $ & $\left\langle R_{s}\right\rangle (nm)$ & $%
\sigma (nm)$ & $\chi ^{2}$ \\ \hline
3.45 to 9.05 & 260.09$\pm $9.81 & 12.66$\pm $19.81 & 1.64 \\ \hline
3.45 to 11.18 & 260.30$\pm $1.49 & 12.30$\pm $3.37 & 1.65 \\ \hline
3.45 to 13.23 & 253.45$\pm $0.69 & 22.80$\pm $0.94 & 2.26 \\ \hline
3.45 to 14.21 & 254.10$\pm $0.15 & 21.94$\pm $0.36 & 2.03 \\ \hline
3.45 to 15.17 & 254.34$\pm $0.12 & 21.47$\pm $0.33 & 2.15 \\ \hline
3.45 to 17.00 & 255.40$\pm $0.10 & 17.32$\pm $0.22 & 11.02 \\ \hline
5.50 to 15.17 & 254.24$\pm $0.15 & 21.95$\pm $0.47 & 2.32 \\ \hline
7.95 to 15.17 & 254.32$\pm $0.16 & 21.56$\pm $0.57 & 2.38 \\ \hline
10.12 to 15.17 & 254.65$\pm $0.10 & 17.81$\pm $0.63 & 0.79 \\ \hline
12.21 to 15.17 & 254.84$\pm $0.16 & 19.33$\pm $0.87 & 0.42 \\ \hline
\end{tabular}
}$
\end{center}

If the reflected light was considered, Eq. \ref{mainre} was used to fit all
data in the full scattering vector range for various factors of reflected
light $b$. The fit results are listed in Table 4.2. The fit results show
that the values of $\chi ^{2}$ are too big, the value of mean static radius $%
\left\langle R_{s}\right\rangle $ is equal to that obtained using Eq. \ref
{mainfit} in the fitting range with the small values of the scattering
vector and the standard deviation changes to small.

\begin{center}
$\overset{\text{Table 4.2 The fit results for }PNIPAM-1\text{ were obtained
using Eq. \ref{mainre}.}}{
\begin{tabular}{|c|c|c|c|}
\hline
$b$ & $\left\langle R_{s}\right\rangle (nm)$ & $\sigma (nm)$ & $\chi ^{2}$
\\ \hline
0.01 & 254.0$\pm $0.3 & 14.4$\pm $0.5 & 194.60 \\ \hline
0.011 & 254.0$\pm $0.3 & 14.6$\pm $0.5 & 168.20 \\ \hline
0.012 & 254.0$\pm $0.3 & 14.7$\pm $0.5 & 149.99 \\ \hline
0.013 & 254.0$\pm $0.2 & 14.8$\pm $0.4 & 139.82 \\ \hline
0.014 & 254.1$\pm $0.2 & 15.0$\pm $0.4 & 137.52 \\ \hline
0.015 & 254.1$\pm $0.2 & 15.1$\pm $0.4 & 142.96 \\ \hline
0.016 & 254.09$\pm $0.07 & 15.2$\pm $0.5 & 155.97 \\ \hline
0.017 & 254.1$\pm $0.3 & 15.4$\pm $0.5 & 176.40 \\ \hline
0.018 & 254.1$\pm $0.3 & 15.5$\pm $0.5 & 204.08 \\ \hline
\end{tabular}
}$
\end{center}

As discussed above, light interference influences the fit results. In order
to eliminate the effects of light interference, the experimental data in the
vicinity of the scattered intensity minimum were neglected. Thus Eq. \ref
{mainre} was used to fit the experimental data in the full scattering vector
range again. The fit values are shown in Table 4.3. The values can be
thought to be consistent with the fit results obtained using Eq. \ref
{mainfit} in the fitting range with the small values of the scattering
vector .

\begin{center}
$\overset{
\begin{array}{c}
\text{Table 4.3 The fit results for }PNIPAM-1\text{ were obtained} \\ 
\text{\ using Eq. \ref{mainre} and neglecting some experimental data.}
\end{array}
}{
\begin{tabular}{|c|c|c|c|}
\hline
$b$ & $\left\langle R_{s}\right\rangle (nm)$ & $\sigma (nm)$ & $\chi ^{2}$
\\ \hline
0.013 & 251.3$\pm $0.6 & 22.17$\pm $0.05 & 79.80 \\ \hline
0.014 & 251.1$\pm $0.6 & 23.3$\pm $0.9 & 58.29 \\ \hline
0.015 & 250.9$\pm $0.6 & 24.4$\pm $0.8 & 44.50 \\ \hline
0.016 & 250.7$\pm $0.5 & 25.4$\pm $0.7 & 37.02 \\ \hline
0.017 & 250.5$\pm $0.6 & 26.4$\pm $0.7 & 36.01 \\ \hline
0.018 & 250.3$\pm $0.6 & 27.24$\pm $0.8 & 41.59 \\ \hline
\end{tabular}
}$
\end{center}

Because 0.45 $\mu m$ filters were used to do dust free for our samples, we
can think that the very big particles do not exist. So the expected values
calculated using Eq. \ref{mainre} and the fit results obtained using Eq. \ref
{mainfit}\ over the fitting range from 0.00345 $nm^{-1}$ to 0.01517 $nm^{-1}$
can be consistent with the experimental data if the number distribution was
corrected. The expected values calculated in three different situations for $%
PNIPAM-1$ are shown in Fig. 4.2. First, the expected results of the incident
light calculated during the full particle size distribution range. Next, a
truncated Gaussian was used, ie. integrated between about the $\left\langle
R_{s}\right\rangle -1.3\sigma $ and $\left\langle R_{s}\right\rangle
+1.3\sigma $ instead of between 1 and 800 nm and the expected values of the
incident light were thus calculated in this truncated Gaussian distribution.
Finally, the integrated range is the same as the second, the expected values
of the incident and the reflected light were calculated with $b$: 0.014.
That the expected values in the third are consistent with the experimental
data show the average scattered intensity is very sensitive to the particle
size distribution.

For the particles with small sizes, the fit results are shown in Table 4.4.
The sample is $PNIPAM-5$. The data were measured at a temperature of 40$%
^{o}C.$ The fit results during the scattering vector range from 0.00345 $%
nm^{-1}$ to 0.02555 $nm^{-1}$ are chosen as the size information obtained
using the SLS technique since the values of $\left\langle R_{s}\right\rangle 
$ and $\sigma $ stabilize. Figure\ 4.3 shows the fit results and the
residuals during the scattering vector range from 0.00345 $nm^{-1}$ to
0.02555 $nm^{-1}$.

\begin{center}
$\overset{\text{Table 4.4 The fit results for }PNIPAM-5\text{ at different
scattering vector ranges and a temperature of 40}^{o}C\text{.}}{
\begin{tabular}{|c|c|c|c|}
\hline
$q\left( 10^{-3}nm^{-1}\right) $ & $\left\langle R_{s}\right\rangle \left(
nm\right) $ & $\sigma \left( nm\right) $ & $\chi ^{2}$ \\ \hline
3.45 to 14.21 & 143.78$\pm $8.34 & 7.32$\pm $14.78 & 2.68 \\ \hline
3.45 to 16.10 & 116.70$\pm $7.35 & 27.01$\pm $3.40 & 2.97 \\ \hline
3.45 to 17.87 & 130.01$\pm $3.54 & 19.45$\pm $2.41 & 2.92 \\ \hline
3.45 to 19.50 & 142.29$\pm $2.33 & 7.26$\pm $4.48 & 4.43 \\ \hline
3.45 to 20.98 & 138.18$\pm $1.50 & 13.47$\pm $1.57 & 3.97 \\ \hline
3.45 to 23.46 & 142.30$\pm $0.57 & 7.97$\pm $1.22 & 3.18 \\ \hline
3.45 to 24.44 & 140.09$\pm $0.46 & 11.59$\pm $0.73 & 3.66 \\ \hline
3.45 to 25.23 & 139.57$\pm $0.41 & 12.33$\pm $0.63 & 3.87 \\ \hline
3.45 to 25.55 & 139.34$\pm $0.31 & 12.36$\pm $0.55 & 5.50 \\ \hline
\end{tabular}
}$
\end{center}

For\ $PNIPAM-5$, due to the values of $qR_{s}$ are small, so the Zimm plot
can be used to obtain the approximative value of the root\ mean square
radius of gyration $\left\langle R_{g}^{2}\right\rangle _{Zimm}^{1/2}$. The
results of the Zimm plot is shown in Fig. 4.4. The value of $\left\langle
R_{g}^{2}\right\rangle _{Zimm}^{1/2}$ is about 115.55 $nm$.

\subsection{Simulated Data Analysis}

In order to conveniently discuss the effects of the reflected light and the
noises, the simulated data have been produced with a Gaussian distribution.

\subsubsection{Simulated data of large particles}

For large particles, both the effects of the reflected light and the noises
must be considered. First, the effects of the reflected light are
considered. The simulated data of the incident light were produced using Eq. 
\ref{mainfit} and the data of the reflected light were obtained using the
following equation

\begin{equation}
\frac{I_{s}}{I_{inc}}=a\frac{4\pi \rho }{3}\frac{b\int_{0}^{\infty
}R_{s}^{6}P\left( q^{\prime },R_{s}\right) G\left( R_{s}\right) dR_{s}}{%
\int_{0}^{\infty }R_{s}^{3}G\left( R_{s}\right) dR_{s}}.
\end{equation}
Then the 1\% statistical noises were added to the simulated data
respectively. Next we will keep the simulated data and only consider the
effects of the reflected light.

The scattered intensity of the reflected light was added to the total
scattered intensities. When the final data of $\frac{I_{s}}{I_{inc}}$ were
obtained, the 3\% random errors were set. The simulated data 1 was produced
when the mean radius was set 267 $nm$ and the standard deviation was 23 $nm$%
. The fit results using Eq. \ref{mainfit} at different scattering vector
ranges are listed in Table 4.5 for the value of $b$ was chosen to be 0.015.
The results show that the fit results' errors decrease and the mean radius
and the standard deviation stabilize when the fitting scattering vector
range is enlarged. Figure 4.5 shows the fit results and the residuals during
the scattering vector range from 0.00345 $nm^{-1}$ to 0.01592 $nm^{-1}$.

\begin{center}
$\overset{\text{Table 4.5 The fit results for the simulated data 1 with b:
0.015 at different scattering vector ranges.}}{
\begin{tabular}{|c|c|c|c|}
\hline
$q\left( 10^{-3}nm^{-1}\right) $ & $\left\langle R_{s}\right\rangle (nm)$ & $%
\sigma (nm)$ & $\chi ^{2}$ \\ \hline
3.45 to 10.97 & 272.52$\pm $1.76 & 12.97$\pm $4.56 & 0.81 \\ \hline
3.45 to 12.01 & 271.18$\pm $1.06 & 16.00$\pm $2.30 & 0.75 \\ \hline
3.45 to 13.02 & 269.41$\pm $0.68 & 19.40$\pm $1.25 & 0.76 \\ \hline
3.45 to 14.02 & 267.45$\pm $0.22 & 22.64$\pm $0.42 & 0.82 \\ \hline
3.45 to 14.98 & 266.95$\pm $0.03 & 23.53$\pm $0.17 & 0.88 \\ \hline
3.45 to 15.92 & 266.96$\pm $0.02 & 23.45$\pm $0.09 & 0.82 \\ \hline
\end{tabular}
}$
\end{center}

Since the Gaussian distribution was used to produce the simulated data, so
Eq. \ref{mainre} can be used to fit the data at the full scattering vector
range. The fit results with the various values of $b$ are listed in Table
4.6. The fit results are consistent with those obtained using Eq. \ref
{mainfit} in the fitting range with the small values of the scattering
vector.

\begin{center}
$\overset{\text{Table\ 4.6 The fit results for the simulated data 1 with }b%
\text{: 0.015 were obtained using Eq. \ref{mainre}.}}{
\begin{tabular}{|c|c|c|c|}
\hline
$b$ & $\left\langle R_{s}\right\rangle (nm)$ & $\sigma (nm)$ & $\chi ^{2}$
\\ \hline
0.012 & 266.62$\pm $0.22 & 22.80$\pm $0.24 & 28.06 \\ \hline
0.013 & 266.67$\pm $0.15 & 22.91$\pm $0.17 & 13.64 \\ \hline
0.014 & 266.72$\pm $0.09 & 23.01$\pm $0.098 & 4.73 \\ \hline
0.015 & 266.758$\pm $0.001 & 23.14$\pm $0.03 & 1.00 \\ \hline
0.016 & 266.83$\pm $0.07 & 23.22$\pm $0.08 & 2.76 \\ \hline
0.017 & 266.89$\pm $0.13 & 23.32$\pm $0.14 & 9.39 \\ \hline
0.018 & 266.95$\pm $0.198 & 23.43$\pm $0.21 & 20.87 \\ \hline
\end{tabular}
}$
\end{center}

The fit results obtained using Eq. \ref{mainfit} in the scattering vector
range from 0.00345 $nm^{-1}$ to 0.01592 $nm^{-1}$ were input Eq. \ref
{mainfit} and Eq. \ref{mainre} to calculate the expected values at the full
scattering vector range with $b$: 0.015, respectively. The results are shown
in Fig. 4.6. The expected results are consistent with the simulated data.

In order to investigate the effects of the reflected light on the fit
results obtained using Eq. \ref{mainfit} in a fitting range with the small
values of the scattering vector. The simulated data were produced for $b$%
=0.0, 0.005, 0.01 and 0.02 respectively. The fit results at the same
scattering vector range from 0.00345 $nm^{-1}$ to 0.01592 $nm^{-1}$ are
listed in Table 4.7. The fit values show that the size information can be
accurately obtained using Eq. \ref{mainfit} in the fitting range with the
small values of the scattering vector and the effects of the reflected light
do not need to be considered.

\begin{center}
$\overset{\text{Table 4.7 The fit results for the simulative data 1 with the
different reflected light.}}{
\begin{tabular}{|c|c|c|c|}
\hline
$b$ & $\left\langle R_{s}\right\rangle (nm)$ & $\sigma (nm)$ & $\chi ^{2}$
\\ \hline
0 & 267.15$\pm $0.02 & 23.103$\pm $0.09 & 0.84 \\ \hline
0.005 & 267.08$\pm $0.02 & 23.22$\pm $0.09 & 0.82 \\ \hline
0.01 & 267.01$\pm $0.02 & 23.42$\pm $0.13 & 0.83 \\ \hline
0.015 & 266.96$\pm $0.02 & 23.45$\pm $0.09 & 0.82 \\ \hline
0.020 & 266.89$\pm $0.02 & 23.56$\pm $0.09 & 0.83 \\ \hline
\end{tabular}
}$
\end{center}

Second, the effects of noises will be considered. The fit results of
simulated data 1 with different noises and the various values of $b$ in the
scattering vector range from 0.00345 $nm^{-1}$ to 0.01592 $nm^{-1}$ are
shown in Table 4.8. The results show that the noises do not influence the
fit values.

\begin{center}
$\overset{\text{Table 4.8 The fit results for the simulated data 1 with the
different noises and reflected light.}}{
\begin{tabular}{|c|c|c|c|}
\hline
& $\left\langle R_{s}\right\rangle (nm)$ & $\sigma (nm)$ & $\chi ^{2}$ \\ 
\hline
0 & 267.141$\pm $0.001 & 23.09$\pm $0.06 & 1.20 \\ \hline
0.005 & 266.91$\pm $0.07 & 23.2$\pm $0.1 & 1.55 \\ \hline
0.01 & 266.82$\pm $0.03 & 23.30$\pm $0.04 & 0.61 \\ \hline
0.015 & 266.95$\pm $0.03 & 23.736$\pm $0.08 & 1.71 \\ \hline
0.02 & 266.82$\pm $0.04 & 23.6$\pm $0.2 & 2.91 \\ \hline
\end{tabular}
}$
\end{center}

\subsubsection{Simulated data of small particles}

For small particles, only the effects of noises need to be considered. The
simulated data 2 was produced when the mean radius was set 90 $nm$ and the
standard deviation was 7 $nm$. The final simulated data were obtained in two
different situations: one is that the noises were not added $\left(
first\right) $ and the other is that the 1\% statistical noises were added $%
\left( \text{second to fifth}\right) $. The fit results are listed in Table
4.9. The values show that the fit results are influenced by noises. Figure
4.7 shows the fit results and residuals for the fifth simulated data of the
simulated data 2.

\begin{center}
$\overset{\text{Table 4.9 The fit results for the simulated data 2 with
different noises.}}{
\begin{tabular}{|c|c|c|c|}
\hline
& $\left\langle R_{s}\right\rangle (nm)$ & $\sigma (nm)$ & $\chi ^{2}$ \\ 
\hline
First & 89.97$\pm $0.08 & 7.02$\pm $0.09 & 0.004 \\ \hline
Second & 87.2$\pm $3.0 & 10.5$\pm $2.5 & 1.41 \\ \hline
Third & 79.0$\pm $3.4 & 15.4$\pm $1.7 & 2.02 \\ \hline
Fourth & 77.5$\pm $2.3 & 16.2$\pm $1.2 & 0.84 \\ \hline
Fifth & 91.2$\pm $1.2 & 4.9$\pm $2.0 & 1.94 \\ \hline
\end{tabular}
}$
\end{center}

If the simulated data 2 with the different noises are put together, as shown
in Fig. 4.8, the differences among the simulated data 2 with the different
noises cannot be distinguished. From the Zimm plot analysis, the root mean
square radius of gyration $\left\langle R_{g}^{2}\right\rangle _{Zimm}^{1/2}$
and the Zimm's molar mass of the particles will be the same. However, due to
the size distribution, the average molar mass of particles $\left\langle
M\right\rangle $ will have large differences for the particles with
different distributions. Figure 4.9 shows the results using the Zimm plot to
fit the third simulated data of the simulated data 2. For the five simulated
data, the fit values of $\left\langle R_{g}^{2}\right\rangle _{Zimm}^{1/2}$
are listed in Table 4.10. If the symbol $k$ is used to represent the
quantity $M_{z}/\left\langle M\right\rangle $. The expected values of $%
\left\langle R_{g}^{2}\right\rangle _{cal}^{1/2}$ and $k$ obtained using
Eqs. \ref{RG} and \ref{Momass} are also shown in Table 4.10.

\begin{center}
$\overset{\text{Table 4.10 Values of }\left\langle R_{g}^{2}\right\rangle
_{Zimm}^{1/2}\text{ , }\left\langle R_{g}^{2}\right\rangle _{cal}^{1/2}\text{
and }k\text{.}}{
\begin{tabular}{|c|c|c|c|}
\hline
& $\left\langle R_{g}^{2}\right\rangle _{Zimm}^{1/2}\left( nm\right) $ & $%
\left\langle R_{g}^{2}\right\rangle _{cal}^{1/2}\left( nm\right) $ & $k$ \\ 
\hline
First & 74.96 & 72.36 & 1.05 \\ \hline
Second & 75.46 & 73.46 & 1.13 \\ \hline
Third & 73.67 & 73.97 & 1.32 \\ \hline
Fourth & 73.12 & 74.17 & 1.36 \\ \hline
Fifth & 74.96 & 71.95 & 1.03 \\ \hline
\end{tabular}
}$
\end{center}

Since the value of $k$ has a strong dependence on the distribution of
particles, the simulated data 3 was produced as the simulated data 2 with a
mean static radius 50 $nm$ and a standard deviation 10 $nm$. The fit results
are listed in Table 4.11. The fit values of $\left\langle
R_{g}^{2}\right\rangle _{Zimm}^{1/2}$, the expected values of $\left\langle
R_{g}^{2}\right\rangle _{cal}^{1/2}$ and $k$ are shown in Table 4.12.

\begin{center}
$\overset{\text{Table 4.11 The fit results for the simulated data 3 with
different noises.}}{
\begin{tabular}{|c|c|c|c|}
\hline
& $\left\langle R_{s}\right\rangle (nm)$ & $\sigma (nm)$ & $\chi ^{2}$ \\ 
\hline
First & 50.5$\pm $0.2 & 9.76$\pm $0.08 & 9.9*10$^{-5}$ \\ \hline
Second & 57.3$\pm $3.2 & 5.1$\pm $3.0 & 0.46 \\ \hline
Third & 42.3$\pm $5.4 & 12.5$\pm $1.9 & 3.39 \\ \hline
Fourth & 58.7$\pm $0.8 & 3.9$\pm $0.9 & 0.30 \\ \hline
Fifth & 58.96$\pm $0.01 & 3.6$\pm $0.2 & 2.57 \\ \hline
\end{tabular}
}$

$\overset{\text{Table 4.12 Values of }\left\langle R_{g}^{2}\right\rangle
_{Zimm}^{1/2}\text{, }\left\langle R_{g}^{2}\right\rangle _{cal}^{1/2}\text{
and }k.}{
\begin{tabular}{|c|c|c|c|}
\hline
& $\left\langle R_{g}^{2}\right\rangle _{Zimm}^{1/2}\left( nm\right) $ & $%
\left\langle R_{g}^{2}\right\rangle _{cal}^{1/2}\left( nm\right) $ & $k$ \\ 
\hline
First & 48.41 & 47.17 & 1.31 \\ \hline
Second & 48.40 & 46.58 & 1.07 \\ \hline
Third & 47.72 & 46.36 & 1.67 \\ \hline
Fourth & 48.47 & 46.74 & 1.04 \\ \hline
Fifth & 46.34 & 46.75 & 1.03 \\ \hline
\end{tabular}
}$
\end{center}

\section{Results and Discussion}

From the analysis of simulated data, for large particles, the reflected
light and the noises do not need to be considered when the size information
is obtained from the SLS\ data in the fitting range with the small values of
the scattering vector. We ever produced the simulated data for the wide
distributions and much larger sizes. The conclusion is the same. For wide
distributions, the mean radius 267 $nm$ and the standard deviation 134 $nm$
were used to produce the simulated data, the fit results during the
scattering vector range 0.00345 $nm^{-1}$ to 0.01498 $nm^{-1}$ are that the
mean radius is 267.5$\pm $0.9 $nm$, the standard deviation is 134.2$\pm $0.5 
$nm$ and $\chi ^{2}$ is 0.91. The expected values calculated inputting the
results in Eqs. \ref{mainfit} and \ref{mainre} respectively are shown in
Fig. 5.1.

For the much larger size, the mean radius 500 $nm$ and the standard
deviation 15 $nm$ were used to produce the simulated data, the fit results
during the scattering vector range from 0.00345 $nm^{-1}$ to 0.00969 $%
nm^{-1} $ are that the mean radius is 500.10$\pm $0.04 $nm$, the standard
deviation is 15.20$\pm $0.04 $nm$ and $\chi ^{2}$ is 0.37. The expected
values calculated inputting the results in Eqs. \ref{mainfit} and \ref
{mainre} respectively are shown in Fig. 5.2.

How the values of $\left\langle R_{g}^{2}\right\rangle _{Zimm}^{1/2}$ can be
obtained\ has been shown. The fit results choosing the different data points
for the simulated data 2 are listed in Table 5.1. The results show that the
values almost keep a constant for the different fitting ranges.

\begin{center}
$\overset{\text{Table 5.1 The values of }\left\langle R_{g}^{2}\right\rangle
_{Zimm}^{1/2}\text{ of the simlated data 2 with the different noises.}}{
\begin{tabular}{|c|c|c|c|c|c|}
\hline
\multicolumn{6}{|c|}{$\left\langle R_{g}^{2}\right\rangle
_{Zimm}^{1/2}\left( nm\right) $} \\ \hline
Fit points & First & Second & Third & Fourth & Fifth \\ \hline
1 to 15 & 74.46 & 70.60 & 73.16 & 74.62 & 78.19 \\ \hline
1 to 20 & 74.96 & 75.50 & 73.67 & 73.13 & 74.96 \\ \hline
1 to 25 & 75.62 & 76.26 & 75.99 & 75.12 & 74.65 \\ \hline
1 to 30 & 76.43 & 77.58 & 77.46 & 77.73 & 75.64 \\ \hline
1 to 35 & 77.28 & 77.76 & 77.61 & 78.07 & 76.48 \\ \hline
\end{tabular}
}$
\end{center}

Since the sizes of PNIPAM microgel particles at high temperatures are small,
the fit values will be influenced by noises, but this method still can give
the better values of $\left\langle R_{g}^{2}\right\rangle _{Zimm}^{1/2}$ and
make us avoid the stringent requirements for the sample quality and the
instrument capability at smaller scattering angles. The fit values of the
four PNIPAM microgel samples at high temperatures are listed in Table 5.2.
All the experimental data of $PNIPAM-5$ and the results obtained using the
Zimm plot in a range with the small values of the scattering vector are
shown in Fig. 5.3. The picture shows that the values of $\left\langle
R_{g}^{2}\right\rangle _{Zimm}^{1/2}$ obtained using a Zimm plot have a
large uncertainty. The value is determined by the chosen data. Even if the
data points that obviously deviate from the linear range were neglected, the
values of $\left\langle R_{g}^{2}\right\rangle _{Zimm}^{1/2}$ still show a
strong dependence on the fit points. For the four PNIPAM microgel samples,
the fit results obtained from a Zimm plot analysis are shown in Table 5.3.

\begin{center}
$\overset{\text{Table 5.2\ The fit results for the four PNIPAM microgel
samples at high temperatures.}}{
\begin{tabular}{|c|c|c|c|}
\hline
Sample $\left( Temperature\right) $ & $\left\langle R_{s}\right\rangle
\left( nm\right) $ & $\sigma \left( nm\right) $ & $\chi ^{2}$ \\ \hline
$PNIPAM-5$ $\left( 40^{o}C\right) $ & 139.3$\pm $0.3 & 12.4$\pm $0.6 & 5.50
\\ \hline
$PNIPAM-2$ $\left( 40^{o}C\right) $ & 114.4$\pm $0.9 & 11.4$\pm $1.1 & 4.34
\\ \hline
$PNIPAM-1$ $\left( 40^{o}C\right) $ & 111.7$\pm $0.9 & 14.8$\pm $0.8 & 2.73
\\ \hline
$PNIPAM-0$ $\left( 40^{o}C\right) $ & 101.7$\pm $1.1 & 8.6$\pm $1.3 & 1.80
\\ \hline
$PNIPAM-0$ $\left( 34^{o}C\right) $ & 93.4$\pm $1.9 & 24.5$\pm $0.9 & 1.56
\\ \hline
\end{tabular}
\ }$
\end{center}

The fit values for the experimental data of the four PNIPAM microgel samples
measured at high temperatures were input to Eqs. \ref{RG} and \ref{Momass}
respectively to obtain the expected values of $\left\langle
R_{g}^{2}\right\rangle _{cal}^{1/2}$ and $k$. The values are listed in Table
5.3.

\begin{center}
$\overset{\text{Table 5.3 Values of }\left\langle R_{g}^{2}\right\rangle
_{cal}^{1/2}\text{ }\left\langle R_{g}^{2}\right\rangle _{Zimm}^{1/2}\text{%
and }k\text{ for the four PNIPAM microgel samples at high temperatures.}}{
\begin{tabular}{|c|c|c|c|}
\hline
Sample $\left( Temperature\right) $ & $\left\langle R_{g}^{2}\right\rangle
_{cal}^{1/2}\left( nm\right) $ & $\left\langle R_{g}^{2}\right\rangle
_{Zimm}^{1/2}\left( nm\right) $ & $k$ \\ \hline
$PNIPAM-5$ $\left( 40^{o}C\right) $ & 113.23 & 113.73 to 122.85 & 1.07 \\ 
\hline
$PNIPAM-2$ $\left( 40^{o}C\right) $ & 94.05 & 89.39 to 125.89 & 1.09 \\ 
\hline
$PNIPAM-1$ $\left( 40^{o}C\right) $ & 95.58 & 88.62 \ to 164.87 & 1.15 \\ 
\hline
$PNIPAM-0$ $\left( 40^{o}C\right) $ & 82.28 & 74.78 to 86.36 & 1.06 \\ \hline
$PNIPAM-0$ $\left( 34^{o}C\right) $ & 97.15 & 102.48 to 113.06 & 1.55 \\ 
\hline
\end{tabular}
\ }$
\end{center}

\section{Conclusion}

The consistency between the theoretical results and SLS data shows that the
size information can be obtained using the non-linear least squares fitting
method and the SLS data contain sensitive size information of particles. For
the large particles, the reflected light and the noises do not influence the
fit results in the range with the small values of the scattering vector. Eq. 
\ref{mainfit} provides a method to measure accurately the particle size
distribution and makes it possible to measure the average molar mass of
large particles if the absolute magnitude of the scattered intensity and
some constants that are related to the instrument and samples are known.

For small size particles, although the fit values are influenced by noises,
it still is a good method to obtain the size information from the SLS data.
It can give a better approximative value of $\left\langle
R_{g}^{2}\right\rangle _{Zimm}^{1/2}$ and avoid the stringent dependences on
the sample quantity and the instrument capability. The molar mass of
particles obtained using the Zimm plot is a better approximative value of
the average molar mass of particles only for the particle systems with very
narrow distributions.

The simple number distributions $G\left( R_{s}\right) $ obtained from the
SLS data are the distributions that people really want to obtain from the
experimental data. They make us avoid the other parameters' effects when the
effects of particle sizes are analyzed.

Fig. 4.1 The experimental and fit results for $PNIPAM-1$ at a temperature of
29$^{o}C$. The circles show the experimental data, the line shows the fit
results and the diamonds show the residuals: $\left( y_{i}-y_{fit}\right)
/\sigma _{i}.$

Fig. 4.2 The experimental and expected results for $PNIPAM-1$. The circles
show the experimental data, the line shows the expected results of the
incident light calculated during the full particle size distribution range,
the dash dot line represents the expected results of the incident light
calculated between about the $\left\langle R_{s}\right\rangle -1.3\sigma $
and $\left\langle R_{s}\right\rangle +1.3\sigma $ and the dot line shows the
expected results of the incident and the reflected light calculated in the
same range as the second with $b$: 0.014$.$ 

Fig. 4.3\ \ The experimental and fit results for $PNIPAM-5$ at a temperature
of 40$^{o}C$. The circles show the experimental data, the line shows the fit
results and the diamonds show the residuals: $\left( y_{i}-y_{fit}\right)
/\sigma _{i}.$

Fig. 4.4 The results of a Zimm plot for $PNIPAM-5$ at a temperature of 40$%
^{o}C$. The circles show the experimental data and the line shows a linear
fit to the plot of $Kc/R_{vv}$ as a function of $q^{2}$.

Fig. 4.5 The simulated and fit results for the simulated data 1 with $b$:
0.015. The circles show the simulated data, the line shows the fit results
and the diamonds show the residuals: $\left( y_{i}-y_{fit}\right) /\sigma
_{i}.$

Fig. 4.6 The simulated and expected results for the simulated data 1 with $b$%
: 0.015. The circles show the simulated data, the line shows the expected
results of the incident light calculated during the full particle size
distribution range and the dot line shows the expected results of the
incident and reflected light calculated during the full particle size
distribution range with $b$: 0.015$.$ 

Fig. 4.7 The simulated and fit results for the fifth simulated data of the
simulated data 2. The circles show the simulated data, the line shows the
fit results and the diamonds show the residuals: $\left(
y_{i}-y_{fit}\right) /\sigma _{i}.$

Fig. 4.8 The simulated data 2 with the different noises.

Fig. 4.9 The results of a Zimm plot for the third simulated data of the
simulated data 2. The circles show the simulated data and the line shows a
linear fit to the plot of $I_{inc}/I_{s}$ as a function of $q^{2}$.

Fig. 5.1 The simulated and expected results for the simulated data. The
circles show the simulated data, the line shows the expected results of the
incident light calculated during the full particle size distribution range
and the dot line shows the expected results of the incident and reflected
light calculated during the full particle size distribution range with $b$:
0.01$.$ 

Fig. 5.2 The simulated and expected results for the simulated data. The
circles show the simulated data, the line shows the expected results of the
incident light calculated during the full particle size distribution range
and the dot line shows the expected results of the incident and reflected
light calculated during the full particle size distribution range with $b$:
0.01$.$ 

Fig. 5.3 The results of the Zimm plot and the experimental data for $PNIPAM-5
$ at a temperature of 40$^{o}C$ in a large scattering vector range. The
circles show the experimental data and the line shows the results of a Zimm
plot obtained in a range with the small values of the scattering vector.

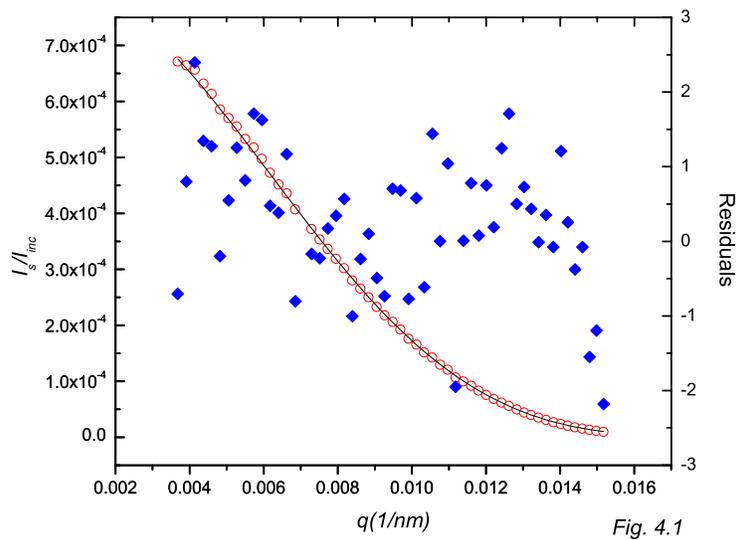
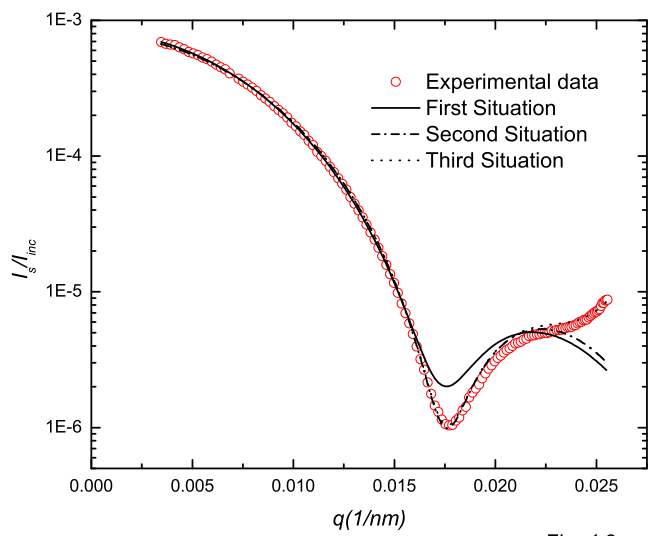
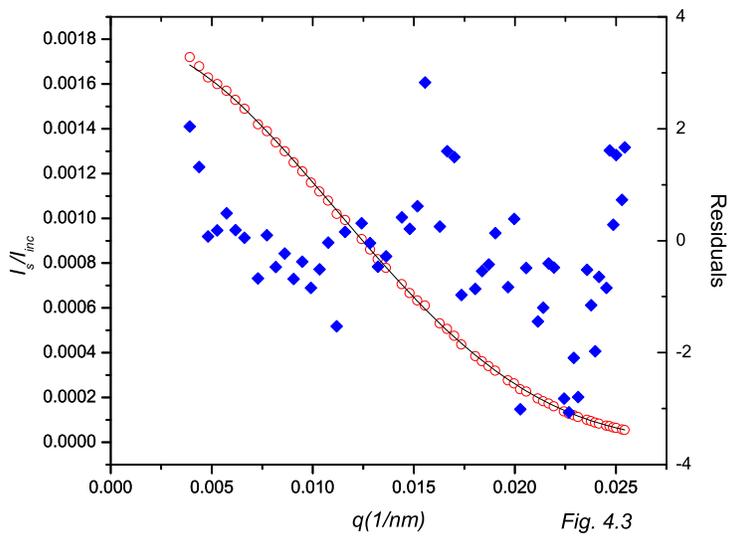
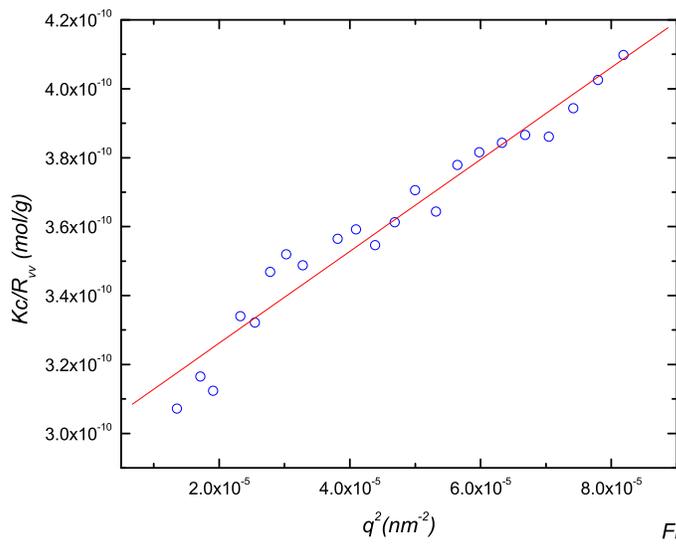
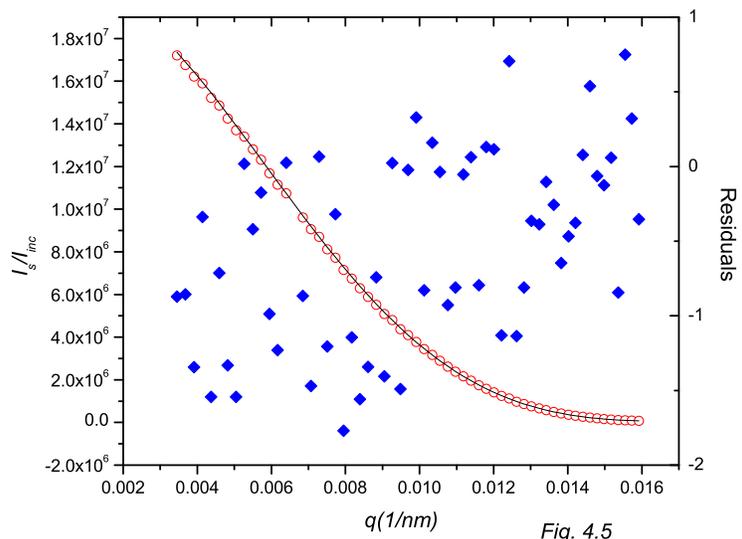
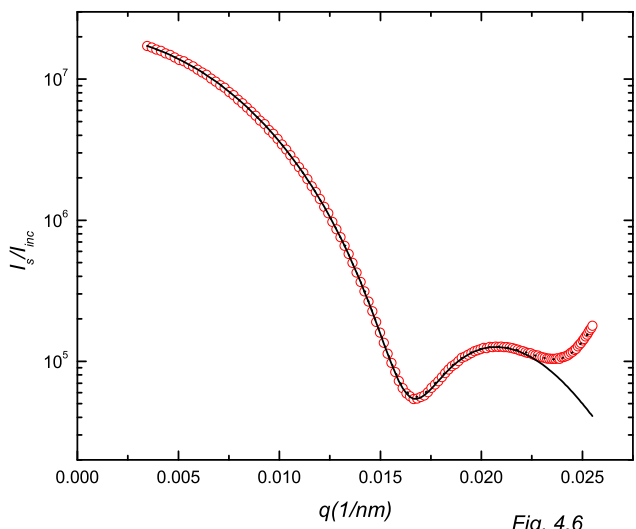
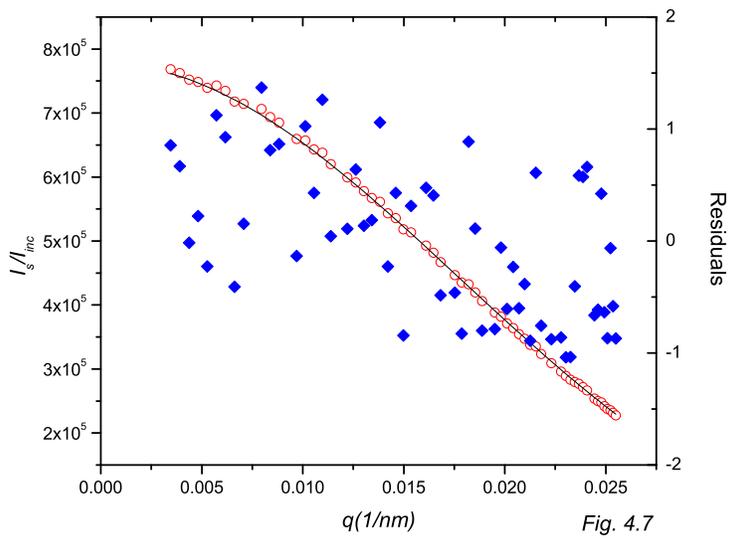
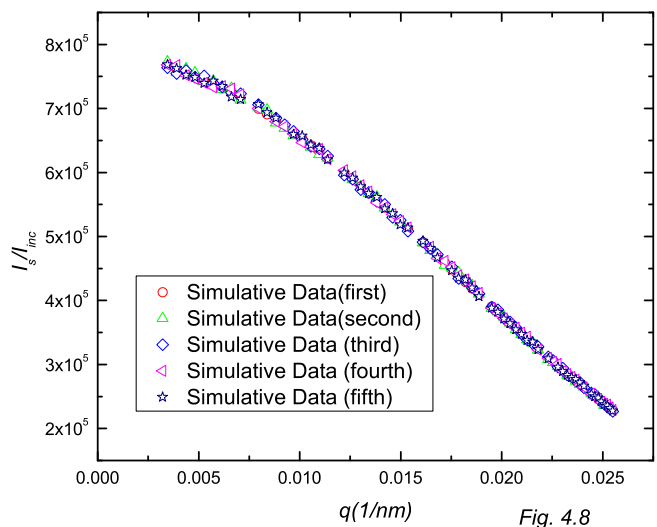
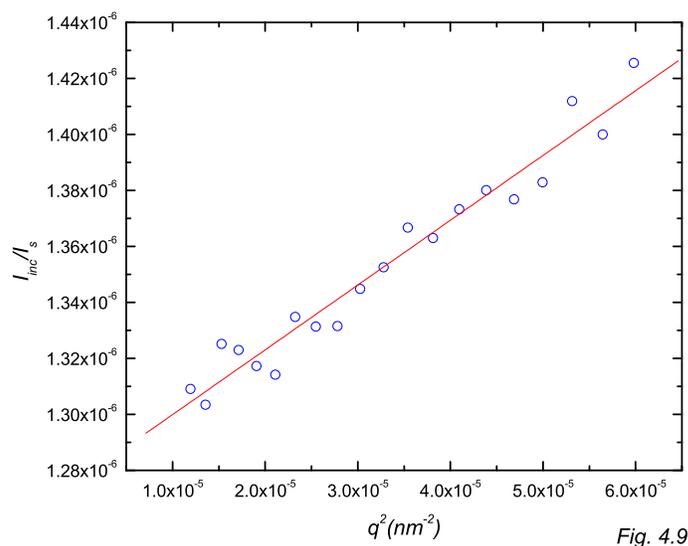
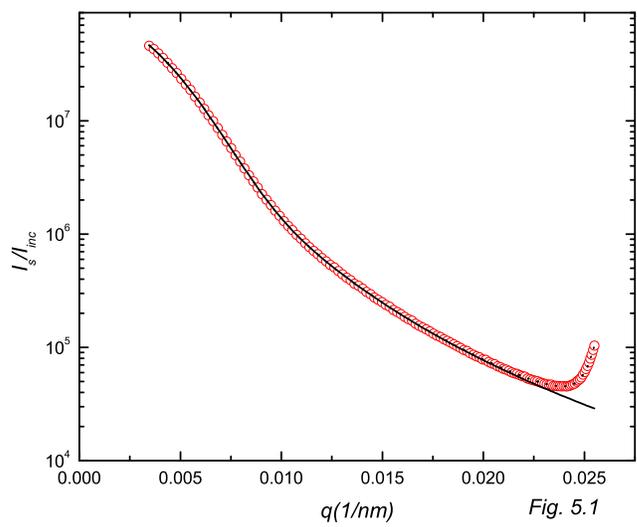
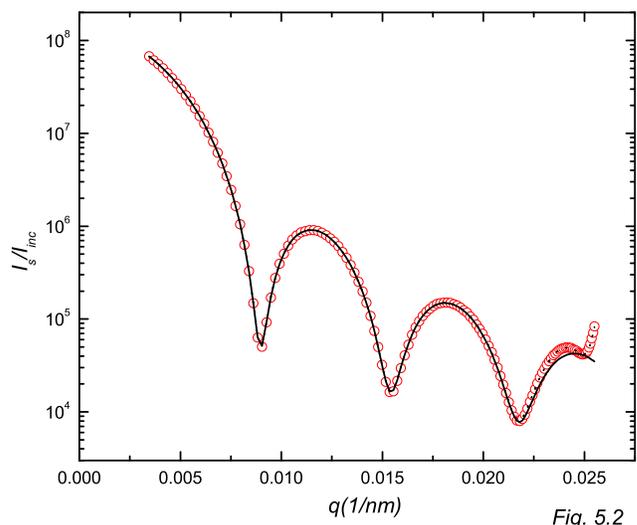
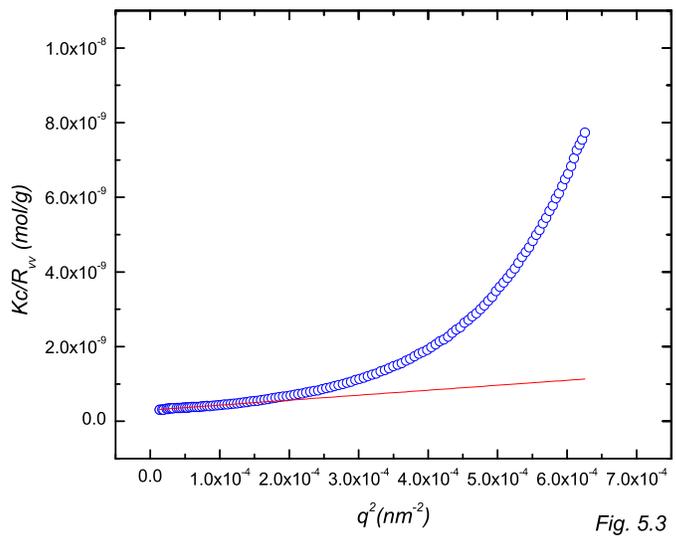
\end{document}